\documentclass[12pt]{article}
\usepackage{graphicx}
\usepackage{amsmath}
\usepackage{amssymb}
\usepackage{latexsym}
\usepackage{hyperref}
\usepackage[sort]{cite}

\newcommand{\be}{\begin{eqnarray}}
\newcommand{\ee}{\end{eqnarray}}
\newcommand{\bc}{\begin{center}}
\newcommand{\ec}{\end{center}}
\newcommand{\bea}{\begin{eqnarray}}
\newcommand{\eea}{\end{eqnarray}}
\newcommand{\ben}{\begin{equation}}

\newcommand{\del}{\partial}

\newcommand{\nn}{\nonumber}

\numberwithin{equation}{section}

\newsavebox{\ns}
\newsavebox{\dbrane}
\newsavebox{\dbshort}

\usepackage{epsfig}
\usepackage{amssymb}

\def\appendix{{\newpage\section*{Appendix}}\let\appendix\section%
        {\setcounter{section}{0}
        \gdef\thesection{\Alph{section}}}\section}

\newcommand\ba{\begin{eqnarray}}
\newcommand\ea{\end{eqnarray}}

\def\Dslash{\,\,{\raise.15ex\hbox{/}\mkern-12mu D}}
\def\Dbarslash{\,\,{\raise.15ex\hbox{/}\mkern-12mu {\bar D}}}
\def\delslash{\,\,{\raise.15ex\hbox{/}\mkern-9mu \partial}}
\def\delbarslash{\,\,{\raise.15ex\hbox{/}\mkern-9mu {\bar\partial}}}
\def\pslash{\,\,{\raise.15ex\hbox{/}\mkern-9mu p}}
\def\calDslash{\,\,{\raise.15ex\hbox{/}\mkern-12mu {\cal D}}}

\newcommand{\hh}{{1\over 2}}

\renewcommand{\ll}{_}
\newcommand{\uu}{^}

\renewcommand{\L}{\Lambda}
\renewcommand{\exp}[1]{{\rm exp}\left( #1 \right)}

\newcommand{\m}{\mu}

\renewcommand{\m}{\mu}

\renewcommand{\t}{\tau}

\newcommand{\g}{\gamma}
\renewcommand{\a}{\alpha}

\newcommand{\e}{\epsilon}

\newcommand{\sqd}{^2}

\renewcommand{\hh}{{1\over 2}}

\newcommand{\eee}[1]{\ba{#1}\ea}
\renewcommand{\th}{\theta}
\renewcommand{\t}{\tau}

\renewcommand{\b}{\beta}

\newcommand{\st}{{}^*}

\newcommand{\pr}{^\prime {}}

\newcommand{\apr}{{\alpha^\prime} {}}

\newcommand{\IZ}{\relax\ifmmode\mathchoice
{\hbox{\cmss Z\kern-.4em Z}}{\hbox{\cmss Z\kern-.4em Z}}
{\lower.9pt\hbox{\cmsss Z\kern-.4em Z}} {\lower1.2pt\hbox{\cmsss
Z\kern-.4em Z}}\else{\cmss Z\kern-.4em Z}\fi} \font\cmss=cmss10
\font\cmsss=cmss10 at 7pt
\newcommand{\inbar}{\,\vrule height1.5ex width.4pt depth0pt}
\newcommand{\IC}{{\relax\hbox{$\inbar\kern-.3em{\rm C}$}}}
\newcommand{\IQ}{{\relax\hbox{$\inbar\kern-.3em{\rm Q}$}}}
\newcommand{\IP}{\relax{\rm I\kern-.18em P}}
\newcommand{\Ione}{{\relax\hbox{$\inbar\kern-.39em{\rm 1}$}}}

\newcommand{\ed}{\dot{e}}

\renewcommand{\L}{\Lambda}
\renewcommand{\pr}{{}^\prime{}}

\newcommand{\pst}{\tilde{\psi}}

\newcommand{\IR}{\relax{\rm I\kern-.18em R}}
\def\blfootnote{\xdef\@thefnmark{}\@footnotetext}

\newcommand{\bm}{\begin{matrix}}

\newcommand{\bbb}{\ba}
\renewcommand{\eee}{\ea}

\def\lrdd{\left(\,  }
\def\rrdd{\, \right)}
\def\lsqq{\left[ \,}
\def\rsqq{\, \right]}

\newcommand{\kket}[1]{\left | {#1} \right \rangle }

\def\bi{\begin{itemize}}
\def\ei{\end{itemize}}

\def\ed{\end{document}}

\def\tb{\bar{\tau}}

\def\mflw{(-1)\uu{\rm F_{\rm L_{\rm w}}}}
\def\mfw{(-1)\uu{\rm F_{\rm w}}}

\def\nev{n_{\rm even}}
\def\nod{n_{\rm odd}}
\def\mfw{(-1)^{F_W}}
\def\mflw{(-1)^{F_{L_W}}}
\def\mfrw{(-1)^{F_{R_W}}}
\def\mfs{(-1)^{F_S}}
\def\mfls{(-1)^{F_{L_S}}}

\renewcommand{\kket}[1]{\left | {{#1}}\right\rangle}
\begin{document}

\begin{titlepage}
\begin{flushright}
arXiv:0705.0980v1
\end{flushright}
\vspace{15 mm}
\begin{center}
  {\Large \bf  Charting the landscape of supercritical string theory}  
\end{center}
\vspace{6 mm}
\begin{center}
{ Simeon Hellerman and Ian Swanson }\\
\vspace{6mm}
{\it School of Natural Sciences, Institute for Advanced Study\\
Princeton, NJ 08540, USA }
\end{center}
\vspace{6 mm}
\begin{center}
{\large Abstract}
\end{center}
\noindent
Special solutions of string theory in supercritical dimensions 
can interpolate in time between theories with different numbers of
spacetime dimensions (via {\it dimension quenching})
and different amounts of worldsheet supersymmetry (via {\it c-duality}).  
These solutions connect supercritical string theories to the more familiar
string duality web in ten dimensions, and provide a precise 
link between supersymmetric and purely bosonic string theories.
Dimension quenching and c-duality appear to be 
natural concepts in string theory, giving
rise to large networks of interconnected theories.
We describe some of these networks in detail and discuss general 
consistency constraints on the types of transitions that arise in 
this framework.  
\vspace{1cm}
\begin{flushleft}
\today
\end{flushleft}
\end{titlepage}
\newpage

\section{\label{intro}Introduction}
Among the major advancements in string theory has been the realization 
that several seemingly distinct perturbative versions of the theory are linked 
by duality, and overwhelming evidence has accumulated that these theories 
collectively emerge from a more fundamental framework known as M theory.
The duality web connects type I, type IIA, type IIB and 
heterotic $SO(32)$ and $E_8\times E_8$ superstring theories.
This is often depicted as a pointed diagram with M theory residing in the
middle, bounded by links among five vertices representing 
individual superstring theories (11-dimensional 
supergravity is sometimes included as an additional vertex).  
This moduli space describes only the duality network 
connecting supersymmetric string theories in a critical number of
spacetime dimensions.  A number of consistent 
string theories possessing either non-critical target space dimensions or 
exhibiting no spacetime supersymmetry (or both) can be included in this picture.

Perturbative string theory in conformal gauge is described by
a $2D$ field theory coupled to $2D$ gravity.
The gravity theory on the string worldsheet possesses an
anomalous Weyl symmetry, and the anomaly can be made to vanish by 
formulating the theory in a specific, critical number of spacetime dimensions, $D_c=10$ for the superstring.
The condition $D=D_{c}$, however, is not the only way to cancel the Weyl 
anomaly.  Consistent string theories 
are known to exist in $D\neq D\ll{c}$ in the presence of a linear dilaton background.
Furthermore, bosonic string theory appears to be another perturbatively self-consistent
physical theory, albeit with an instability.  
Understanding the relationship between both 
noncritical and bosonic string theories
and the more familiar incarnations of critical superstring theory has 
been a longstanding problem.

The solutions we present here solve this problem conclusively.
In fact, we can classify
the complete set of such solutions under certain conditions 
(though more possibilities may be allowed 
when these conditions are relaxed).
The dynamics of our models describe a domain wall moving to the left
at the speed of light, separating two phases described by
distinct string theories. 
The right side of the wall describes a string theory with
lower potential energy.  The configuration as a whole can be thought
of as the late time limit of an \it expanding bubble \rm
of a lower-energy vacuum.  Transitions 
to the new vacuum can be interpreted as
a renormalization group flow in the worldsheet theory, dressed
with an exponential of $X\uu +$ to make the relevant perturbation
strictly scale invariant.

In some instances, the dynamics interpolate between string 
theories in different numbers of 
spacetime dimensions: in passing across the domain 
wall of the expanding bubble, the number of spatial dimensions 
is reduced dynamically.  To simplify the exposition,
we collectively refer to processes in this category as {\it dimension quenching}.
In the examples we study, the stable
endpoint of successive stages of
dimension quenching is string theory in either the critical dimension (for
type II theories) or in two dimensions (for bosonic and type 0 theories). 

We also describe transitions that 
connect type 0 superstring theory
dynamically with purely 
bosonic string theory.  These solutions
constitute precise dualities, 
unique in that
they are realized {\it cosmologically}, rather than formally,
or as adiabatic motion along moduli space.
As with dimension-quenching transitions, 
time evolution in the target space equates to an
RG flow on the worldsheet.  
For this reason, and because they involve a transfer of central charge 
contribution form various sources, we dub these relations {\it c-dualities}.

The intent of this paper is to present a global atlas of
connections between various consistent string theories that arise from the above processes.
The resulting networks, while not exhaustive, 
demonstrate the rich interconnectedness of string theory.
Detailed expositions of dimension quenching and c-duality
can be found in references \cite{previous,previous2,previous3}, 
and in forthcoming work.  


\section{\label{dim}Dimension quenching}
The solutions we study describe a reduction in
the matter central charge on the worldsheet as a function of
light-cone time $X\uu +$, triggered by 
a nonzero tachyon expectation value.
The initial matter central charge is
equal to the number of spacetime dimensions $D$.  There is
a timelike dilaton dependence $\Phi = - q X\uu 0$,
which compensates the matter central charge excess if we set
$q = \sqrt{{D - 26}\over{6\apr}}$ for the bosonic string
or $q = \sqrt{{D - 10}\over{4\apr}}$ for
the superstring.

The worldsheet
dynamics are exactly solvable at the quantum level, despite the fact that 
the underlying $2D$ theories are fully interacting. 
(Quantum corrections are either absent, or are exact 
at one-loop order in perturbation theory.)   
The on-shell condition for a generic tachyon 
vertex operator ${\cal T}(X)$ 
in the linear dilaton background is
\be
\del^2 {\cal T} - 2 V^\mu \del_\mu {\cal T} + \frac{4}{\alpha'}{\cal T} = 0\ ,
\label{onshell}
\ee
where $V^\mu$ is the dilaton gradient.  The second term in 
Eqn.~(\ref{onshell}) signifies 
an anomalous dimension, while the third term represents a 
non-linearly realized contribution to the scaling dimension.  
Non-infinitesimal vertex operators satisfying 
Eqn.~(\ref{onshell}) will generally lead to complicated theories,
since multiple insertions of ${\cal T}(X)$ will become singular
when they approach one another.

There are special choices of ${\cal T}(X)$ for which the worldsheet
theory is well-defined and conformal to all orders in perturbation theory
(and nonperturbatively) in $\alpha'$.
Defining the lightcone frame
$X^\pm \equiv (X^0 \pm X^1) / \sqrt{2}$,
one sees that the exponential
$\exp{\b X\uu +}$ is non-singular in the vicinity of 
another copy of itself.  The conformal properties 
of vertex operators involving  
$\exp{\b X\uu +}$ are therefore completely well-behaved.  
If we choose $\b = {{2\sqrt{2}}
\over q}$, this operator has weight $(1,1)$ and zero
\it anomalous \rm dimension (that is, no piece of the dimension
quadratic in the exponent).  The vanishing
of the anomalous dimension makes it possible to construct
Lagrangian perturbations that deform the free theory while
preserving conformal invariance exactly.

For a profile of the 
form ${\cal T} \sim \exp{\b X\uu +}$,
the tachyon perturbation will increase exponentially
into the future in a lightlike direction.
The dilaton rolls to weak coupling,
and we can arrange for tachyon condensation to occur long after
the big bang.  This means that any effects associated with the 
strongly-coupled region of the cosmology (located in the far past) 
are washed out by the subsequent expansion of the universe.

The tachyon couples to the worldsheet according to 
the Lagrangian
\be
{\cal L} = -\frac{1}{2\pi\alpha'} 
	\left(
	\del_0 X^+ \del_0 X^- - \del_1 X^+ \del_1 X^-
	+ \alpha' \mu^2 \exp{\beta X^+} 
	\right)\ , 
\label{lag1}
\ee
where $\del_{0,1}$ are derivatives along the worldsheet.
This $2D$ field theory is remarkably simple and exactly conformal, 
even at the quantum level.  The potential barrier
accelerates string states to the left,
and their speed rapidly approaches the speed of light.   
Rather than describing a transition
between two conventional string theories,
this simple model connects a timelike linear dilaton
background to a `nothing state,' into which 
no excitation, including the graviton, can propagate.

We can also introduce some dependence on a third 
coordinate $X_2$:
\be
{\cal T} = \mu_0^2 e^{\beta X^+} 
	- \mu_k^2 \cos (k X\ll 2) e^{\beta_k X^+}\ ,
\ee
where
\be
\b\ll k = {{\sqrt{2}}\over{q}} \lrdd {2\over{\apr}} - \hh k\sqd \rrdd\ .
\ee
One obtains a marginal perturbation that is Gaussian 
in $X_2$ in the 
long-wavelength limit $k \to 0$, 
and the resulting theory is exactly solvable.  
Classically, the tachyon becomes large in the future, 
and the theory acquires a mass term for the coordinate $X_2$,
which decreases exponentially:
\be
{\cal T}(X^ +, X_2)
 &=& {{\mu^2}\over{2  \alpha'}}
 \exp{\beta X^+} : X_2^2:
+ {\cal T}_0 (X^+)\ ,
\nn\\
&&
\nn\\
{\cal T}_0 (X^+) &=&  {{\mu^2  X^+}\over
{\alpha' q \sqrt{2}}} \exp {\beta X^+} + 
\mu^{\prime 2} \exp{\beta X^+} \ .
\label{dimq}
\ee

Like the simple case described by the Lagrangian in 
Eqn.~(\ref{lag1}), this theory
describes a domain bubble expanding 
at the speed of light. We can consider two 
basic categories of string trajectories in this background.  
We characterize one set of trajectories as generic, in that they 
carry energy in the $X_2$ direction.  
Upon encountering the domain bubble, 
the $X_2$ field becomes frozen into a state of nonzero 
excitation and is pushed out along
the wall to $X\ll 1 \to -\infty$ at late times (similar to string states in the 
vicinity of the bubble of nothing). 

In addition, there is a special set of trajectories carrying no energy
in the modes of the $X_2$ field.  These states are allowed to propagate 
through the domain wall and into the bubble interior.   
(One might think of the interface as a {\it domain filter}.)
The amount of dynamical matter on the worldsheet therefore
decreases dynamically as a function of $X\uu +$. 
The $X_2$ dimension is quenched for large $X\uu +$, and the
region remaining at late times inside the tachyon condensate 
is a $D-1$ dimensional theory.

\begin{figure}[htp]
\begin{center}
\includegraphics[width=2.6in,height=1.4in,angle=0]{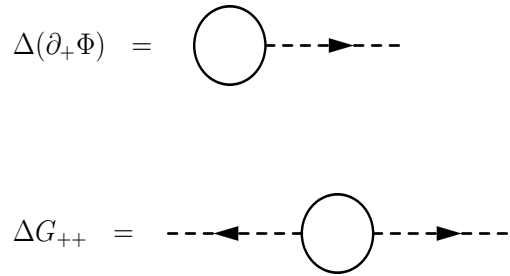}
\caption{\footnotesize
Feynman diagrams contributing to the
renormalizations of the dilaton and metric.  The massive $X^2$ field
(solid lines) propagates in the loops, while the massless fields
(dashed lines) have oriented propagators.  Quantum corrections 
terminate at one-loop order in perturbation theory. }
\label{renorm}
\end{center}
\end{figure}

The contribution to the central charge from the dilaton sector of the theory
is determined by the dilaton gradient and the string-frame metric.
Both of these objects
shift by finite amounts during the dimension-quenching 
transition.  From the worldsheet point 
of view, this is a one-loop renormalization of effective couplings arising 
when the $X\ll 2$ field is integrated out.
From the spacetime perspective, the effect
is a backreaction of the tachyon onto the dilaton and string-frame metric.  
The result is
that the dilaton contribution to the central charge increases by one unit, 
compensating the loss of the $X\ll 2$ degree of freedom.  

In fact, all quantum corrections in this theory are saturated
at one-loop order in perturbation theory, with $X_2$ fields propagating in
the loop. (The $X\uu \pm$ fields have oriented propagators; see \cite{previous2}
for details.)
It is therefore possible to calculate this renormalization 
exactly.   
Most corrections coming from integrating out $X_2$ vanish in the 
$X^+ \to \infty$ limit,
apart from contributions from the effective 
tachyon, the dilaton and the string-frame metric.  
The effective tachyon can be fine-tuned to zero in the future,  
and what remains are corrections to the metric and dilaton 
gradient arising from the one-loop Feynman diagrams
depicted in Fig.~\ref{renorm}.
Taking these contributions into account, the dilaton central charge 
is shifted up by one unit, and the total central charge of the theory 
remains constant.

There are also generalizations of these solutions 
connecting type 0, type II and stable or unstable heterotic string
theories in diverse dimensions.  Here we will focus on connections
among type 0 and type II theories.  (Transitions among
heterotic theories give rise to a web of 
connections even more intricate than the ones we describe
in this article.)  In mapping out the various possibilities, it
is helpful to sort the allowed transitions into the following categories: 
\bi{\item{ {\it Stable transitions}, in which no perturbation
of the solution can destroy or alter the final state qualitatively;}
\item{ {\it Natural transitions},
in which no instability can destabilize
the solution without breaking additional symmetry;}
\item{ {\it Tuned transitions}, in which the initial conditions of
an unstable mode must be fine-tuned to preserve the
qualitative nature of the final state}.}
\ei
Under this classification scheme, transitions among bosonic string
theories in different numbers of dimensions are tuned transitions.  In
particular, the constant $\mu\uu{\prime 2}$ in Eqn.~(\ref{dimq})
must be tuned to
a particular value\footnote{In general, this value will be regulator-dependent.} 
to set the bosonic string tachyon to zero in the
far future of the lower-dimensional final state.

\paragraph{Tuned transitions between type 0 theories with $\Delta D =1$}
\ \\
Transitions between
type 0 string theories in even numbers of dimensions
are natural by virtue of a discrete R-parity
$g\ll L$ that acts with a $+1$ on $G$ and a
$-1$ on $\tilde{G}$ (and, possibly, with a $-1$ on 
some number of coordinates).
However, type 0 string theories in odd dimensions preserve no 
such symmetry (see, e.g.,~\cite{previous2} for a discussion of this
point).  Transitions between
even- and odd-dimensional type 0 theories are therefore tuned
transitions.  Here we consider the case for which the dimension
of spacetime decreases by $\Delta D = 1$.

The tachyon ${\cal T}$ couples to the worldsheet
as a superpotential:\footnote{By `superpotential' we mean an integrand of
a {\it full \rm} (1,1) superspace integral
whose integrand contains only undifferentiated scalars, and
no fermions.  It is the term which contributes to the bosonic
potential and Yukawa couplings.
Note that the superpotential in a $(1,1)$ theory
is not protected from perturbative renormalization.}
\bbb
{\Delta {\cal L}} = \frac{i}{2\pi} \int d\theta\ll +
d\theta \ll - :{\cal T}(X): \ ,
\eee
with
\bbb
:{\cal T}: = ~\exp{\b X\uu +} \lrdd
\frac{\m^2}{2\alpha'} : X\ll 2\sqd: + \frac{\mu^2}{\alpha' q \sqrt{2}}
X\uu +  +  {\m\pr}^2  \rrdd \ .
\eee
The quantity $\m\pr$ is a regulator-dependent coefficient that 
must be tuned to make the effective superpotential
vanish in the limit $X\uu + \to \infty$.
Unlike the even-dimensional type 0 theories, which come in 
two varieties (0A and 0B), there is only one kind of odd-dimensional
type 0 theory.
Starting from odd-dimensional type 0, the sign of $\m$
determines whether one reaches type 0A or type 0B string
theory in $D-1$ dimensions as a final state.  After defining
the odd-dimensional GSO projection with a particular phase
($(-1)\uu{F_W} = \pm i$) in the Ramond/Ramond sectors,
altering the sign of $\m$ changes the ground state of the
$\psi\ll 2, \pst\ll 2$ fermions, and thereby reverses the
effective GSO projection for the $D-1$ remaining Majorana fermions.

\begin{figure}[htp]
\begin{center}
\includegraphics[width=3.8in,height=6.4in,angle=0]{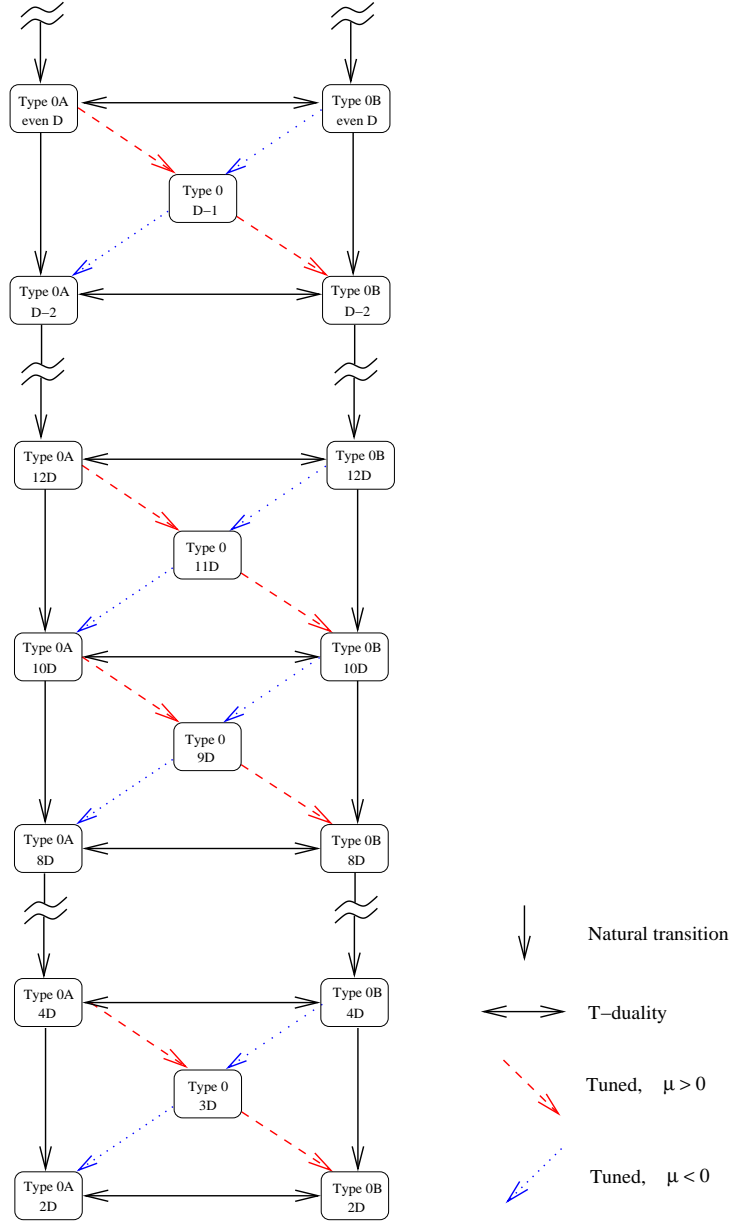}
\caption{\footnotesize
The dimension quenching transitions in type 0 string theory
define a semi-infinite lattice of connected theories.  The
diagonal lines are tuned transitions 
that reduce the spacetime dimensionality by one  
(right-pointing downward
arrows have $\m > 0$, and left-pointing
downward arrows have $\m < 0$).
The vertical lines are
natural transitions reducing the spacetime dimensionality by
two.  The horizontal lines do not represent dynamical transitions;
rather, they represent the standard connection between type
0A and 0B string theory by
T-duality or orbifolding by left-moving spacetime fermion number
$\mfls$ (the lowest horizontal arrow represents orbifolding, or
alternatively a thermal T-duality).  
The lowest point on the diagram represents two-dimensional
string theory of the kind described by the $\hat{c} = 1$ matrix
model.  }
\label{0fig}
\end{center}
\end{figure}

These tuned transitions are represented by the
diagonal lines in Fig.~\ref{0fig}, with right-pointing downward
(dashed) arrows representing transitions with $\m > 0$ and left-pointing
downward (dotted) arrows representing transitions with $\m < 0$.
The tuned transitions can be defined between type 0 linear
dilaton backgrounds in any number of dimensions, 
with any value of $\Delta D$, as long as the final state
is at least two-dimensional.

\paragraph{Natural transitions between even-dimensional type 0 states}
\ \\
When the number of spacetime dimensions reduces by
an even number ($\Delta D = 2K$, $K\in \IZ$) during a transition,
it is possible for the flow to be natural,
in the sense that a discrete symmetry $g\ll L$ can preserve the
qualitative nature of the final state and prevent the generation of
an effective superpotential.  For instance, the tachyon profile
\bbb
{\cal T} = \frac{\mu^2}{2\alpha'} \exp{\b X\uu +} 
~X\ll 2 X\ll 3
\eee
reduces the number of spacetime dimensions by $\Delta D = 2$
while preserving the chiral R-parity $g\ll L$, which acts according to
\be
g_L: \quad 
X\ll 2 \to X\ll 2\ , 
\quad  
X\ll 3 \to - X\ll 3\ ,
\quad 
\th\ll + \to \th\ll + \ ,
\quad 
\th\ll - \to - \th\ll -\ .
\ee
Indeed, $g\ll L$ constitutes a discrete R-symmetry, which
forbids the generation of a superpotential
in the effective theory describing the $D-2$ embedding coordinates
and their superpartners.  These natural transitions are represented 
in Fig.~\ref{0fig} by downward solid arrows.

More generally, the tachyon profile
\bbb
{\cal T} = \frac{\mu^2}{2\alpha'} \exp{\b X\uu +} \lrdd
~X\ll 2 X\ll 3 + X\ll 4 X\ll 5 + \cdots + X\ll {2K}
X\ll{2K+1} \rrdd
\eee
gives rise to a natural transition reducing the spacetime dimension
by $2K$, preserving the discrete global symmetry
\bbb
\th\ll \pm\to \pm\th\ll \pm \ ,
\hskip .75in (X\ll 3, X\ll 5,\cdots , X\ll {2K+1}) \to
- (X\ll 3, X\ll 5,\cdots , X\ll {2K+1}) \ .
\eee
This symmetry
forbids the generation of an effective worldsheet superpotential
for the string propagating in $D-2K$ dimensions.

\paragraph{Stable transitions from type 0 to type II}
\ \\
The natural transitions described above can be converted to
stable transitions if one orbifolds by 
a discrete $R$-symmetry, such as $g\ll L$.  Modular invariance
constrains the conditions under which one can orbifold by
such a symmetry.  A sufficient condition is that the
spacetime dimension be even and
the number $K$ of orbifolded dimensions be equal to 
$\hh(D-10)$.  In this case, the GSO projection 
inherited by the theory in the future
is the chiral GSO projection of the type II
string worldsheet.
As a result, one can condense the tachyon without 
fine tuning initial conditions to recover a stable,
supersymmetric background in the distant future.
Since all relevant operators that could alter the nature of the final state
have been projected out by the action of $g\ll L$, the final
state is stable against all gauge-invariant perturbations.

In the limit $X\uu + \to  \infty$, the massive string modes of
the $10D$ theory have vanishing expectation
value, the dilaton is lightlike and rolling to weak coupling, and the
background preserves half the supersymmetries of type II 
string dynamics.
Starting from an initial state with $D > 10$,
the final state is a half-BPS vacuum of
critical superstring theory.  
This establishes that 
supercritical string theories are connected by 
dimension quenching to the standard duality web of 
critical superstring theory.

\begin{figure}[htp]
\begin{center}
\includegraphics[width=4.2in,height=5.4in,angle=0]{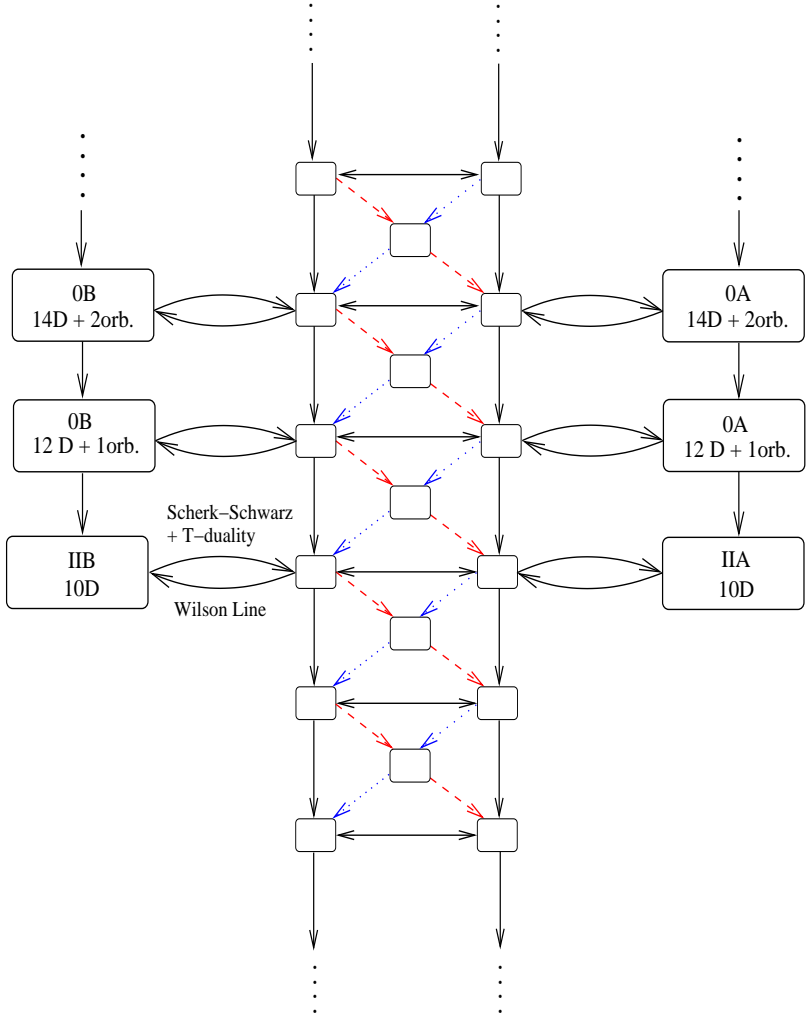}
\caption{\footnotesize
Natural dimension-quenching transitions of orbifolded type 0 string theories 
terminate with type II string theory in the critical dimension ($D=10$).  
These hierarchies are connected laterally
to the type 0 series by Scherk-Schwarz compactification and T-duality, 
or by a discrete Wilson line construction \cite{wilsonization}. 
The type 0 hierarchy in the middle is labeled according to the legend in
Fig.~\ref{0fig}.}  
\label{0IIfig}
\end{center}
\end{figure}

The stable transitions to $10D$ type II string theory are
represented by the vertical arrows on the left- and right-hand
ladders of Fig.~\ref{0IIfig}.
As noted, these transitions are related to the
natural transitions by orbifolding:  These connections are represented
in Fig.~\ref{0IIfig} by the horizontal arrows connecting unorbifolded 
type 0 theories with ladders representing transitions from type 0
orbifolds down to type II in the critical dimension.  The inverse 
connection can be formed by a particular discrete Wilson line construction
of the type described in \cite{wilsonization}.

\paragraph{Relation to timelike tachyon condensation}
\ \\
Any dimension-changing exact solution of the kind
described above can be deformed to make the tachyon
gradient timelike rather than lightlike.  We may
embed both possibilities in a family of theories, where the
exponential dependence of the tachyon is $\exp{B\ll 0 X\uu 0 + B\ll 1 X\uu 1}$,
with $B\ll 1$ varying such that the linearized equations of motion are still satisfied. 

Specifically,
we can vary $B\ll 1$ continuously from $\b / \sqrt{2}$ to $0$, 
taking
\be
B\ll 0 = -q + \left ( B\ll 1 \sqd + q\sqd + |m\sqd\ll{\rm tach.}| \right ) \uu{\hh} \ .
\ee
 The anomalous
dimension of the vertex operator is simply 
\be
\e\equiv {{\apr}\over 2}   \lrdd B\ll 0\sqd - B\ll 1\sqd \rrdd \ .
\ee 
The properties of the tree-level string theory depend on $\Delta D$
(i.e.,~the number of coordinates being eliminated during the transition) 
and on $\e$.  
Fixing $\Delta D$, there are two limits in which $\e\to 0$ and the worldsheet
CFT is solvable.  The first is the case we consider here, in which
$B\ll 0 = B\ll 1$.  The second is the case in which $B\ll 1 = 0$ identically,
and $D\to\infty$ \cite{hellermantwo,freedman}. 
This second limit preserves the full set of symmetries
of a spatially flat FRW cosmology.  However, it cannot describe the return to a
supersymmetric vacuum, since $D\ll{\rm final} = D - \Delta D \gg 10$.  
It has been
proposed \cite{hellermanone,riemann} that solutions may exist with 
$B\ll 1 = 0$ and $D\ll{\rm final} = 
D\ll{c}$, in which 
the theory preserves FRW symmetry and the final state lies in the critical dimension.  
At present, the existence of such solutions remains conjectural, and requires
the absence of a phase transition between $\e = 0$ and $\e$ of
order one.  Another tractable limit of such models is $B\ll 1 = 0$ and $D,~\Delta D\to
\infty$ with $\e \, \Delta D$
held fixed.  In this regime, the worldsheet theory can be analyzed using the techniques
of large-$N$ vector-like models, 
with $\e \, \Delta D$ playing the role of the $2D$ 't~Hooft coupling
$g\sqd N$ \cite{grossneveu,largenreview}.

\paragraph{Relation to 'Dimensional Duality'}
\ \\
Recently, a series of papers \cite{riemann,them} appeared describing 
dynamical transitions between certain timelike linear dilaton
backgrounds in $D>D_c$ as initial states, and final-state backgrounds
on negatively curved spaces in $D=D_c$.  The initial states
of \cite{them} are type 0 string theories with
timelike linear dilaton, running to weak coupling in the future.
The models, parametrized by an integer $h$, have initial-state
geometries described by 7-dimensional flat spatial slices,
a time direction, a 2h-dimensional torus, and $2h-2$ real fiber
directions, which we will call $Y\uu A$.  The $Y$ directions
are fibered over the $2h$-torus, as well as orbfolded by a
$\IZ\ll 2$ symmetry which acts as $Y\uu A \to - Y\uu A$,
as well as with a $-$ sign on the $\tilde{G}$ supercurrent and a
$+$ sign on the $G$ supercurrent.  Indeed, the models of 
\cite{riemann, them} are special cases of the type 0/type II
transitions described
in \cite{previous2}, with the tachyon taken to be timelike
instead of lightlike.
(Order one $\alpha^\prime$ corrections contributed by the
$X\uu 0$ sector must be assumed to
cause no qualitative difference to the dynamics, as in
the approach of \cite{hellermanone}.)

The total number of spacetime dimensions 
in the initial state is $D = 4h+6$.  
The number of orbifolded dimensions is $K = 2h-2$.
As for the natural transition models described earlier in this
section, the number $K$ of orbifolded dimensions is equal
to $\hh (D - 10)$.  In section \ref{classify}, we will
show that under broad conditions,
the number $K$ of orbifolded dimensions
in the initial state must be equal to $\hh(D - 10)$ for 
\it any \rm model of perturbative tachyon condensation
whose endpoint is the type II superstring.  The models of
\cite{them} constitute a nontrivial example of the
classification theorem described in sec. \ref{classify}
for dimension-reducing dynamical transitions with 
critical type II final state.

\section{\label{tachyon}c-duality}
One surprising outcome arises when considering type 0 string theory 
with no orbifold singularities in a flat, timelike linear 
dilaton background.  We will focus first on cases for which 
the tachyon perturbation has no dependence at all on 
the directions transverse to the lightcone.

The type 0 tachyon couples to 
the string worldsheet as a $(1,1)$ superpotential.
The worldsheet theory is thus 
deformed by the following interaction Lagrangian:
\be
{\cal L}_{\rm int.} = -\frac{\alpha'}{8\pi} G^{\mu\nu}
	\del_\mu {\cal T} \del_\nu {\cal T}
	+ \frac{i\alpha'}{4\pi}\del_\mu\del_\nu {\cal T}
	\tilde\psi^\mu \psi^\nu\ ,
\label{type0}
\ee
where $\psi$ and $\tilde\psi$ are right- and left-moving worldsheet
fermions.  In addition, the supersymmetry algebra is modified by the 
following $F$-term
\be
F\uu - = \{Q_- , \psi^M\} &=& - \{Q_+ ,\tilde\psi^M\} = - 
\sqrt{{\alpha'}\over 8}
G^{MN} \del_N{\cal T} \ .
\ee
When the tachyon is lightlike, the worldsheet potential 
vanishes, but the Yukawa coupling (the second term in
Eqn.~(\ref{type0})), the $F$-terms and
the interaction terms all grow as $X^+ \to \infty$. 
Despite this fact, the fermion mass matrix remains nilpotent,
and the physical frequencies of the fermion modes do not grow.
With no worldsheet potential, no states are 
expelled from the interior of the domain bubble, and the number
of spacetime dimensions does not change.  The
presence of the $F$-term, however, indicates that the worldsheet supersymmetry 
is spontaneously broken as $X^+ \to \infty$. 

Although the worldsheet theory is strongly interacting in its
original variables at large $X\uu +$, we 
can describe the physical content precisely
by invoking a series of canonical transformations.  The starting point 
is to exchange the lightcone fermions $\psi\uu\pm$ for a $bc$ ghost system 
with weights ${3\over 2}$ and $-\hh$, denoted by $b\ll 1, c\ll 1$ 
(and similarly for left-movers):
\be
\psi^+ &=& 2 c^\prime\ll 1 - M^{-1} \tilde{b}\ll 1
+ 2\beta (\del_ + X^ +) c\ll 1 \ ,
\nn\\
\tilde\psi^ + &=&  - 2  \tilde{c}^\prime\ll 1  + M^ {-1} b\ll 1  
+ 2 \beta (\del_ - X^ +) \tilde{c}\ll 1 \ ,
\nn\\
\psi^ - &=& M \tilde{c}\ll 1\ ,
\nn\\
\tilde\psi^ - &=& - M c\ll 1 \ ,
\ee
where $M \equiv \mu \, \exp{\beta X^+}$.
In this new set of variables the Lagrangian consists of a free theory
plus a perturbation proportional to $\exp{- \b X\uu +}$, 
which becomes vanishingly small 
deep inside the region of large tachyon condensate.

A subsequent series of canonical transformations can be invoked
that render the stress tensor and supercurrent in the
following form \cite{previous3}:
\be
T    &=& -\frac{3i}{2}\del_+ c\ll 1 \, b\ll 1 - \frac{i}{2}c\ll 1\, \del_+ b\ll 1
	+ \frac{i}{2} \del_+ (c\ll 1 \, \del_+^2 c\ll 1) + T_{\rm matter}\ ,
\nn\\
&&
\nn\\
G &=& b\ll 1 + i \del_+ c\ll 1 \, b\ll 1 \, c\ll 1 - c\ll 1\, T_{\rm matter} 
	- \frac{5}{2} \del_+^2 c\ll 1\ .
\ee
At this stage, the supersymmetry is completely nonlinearly 
realized.  In this form, the $D$-dimensional
theory is a free worldsheet theory with a $bc$ ghost system, 
$D$ free scalars and $D-2$ free fermions (with appropriate
left- or right-moving counterparts).  The central charge
receives a contribution of 26 from the matter-dilaton system
(the dilaton gradient is again renormalized), and $-11$
from the $bc$ ghosts.  The theory therefore exhibits the correct
central charge for the worldsheet of an RNS superstring in conformal 
gauge.

In fact, the system at late times
fits into a construction due to Berkovits 
and Vafa \cite{bv}, in which the bosonic string is embedded
in the solution space of the superstring (i.e.,~it 
is a formal rewriting of the bosonic string as a 
superstring).  This is just a 
manifestation of the fact that to any theory one may add
any amount of nonlinearly realized local symmetry as a redundancy of description.
From the perspective of the bosonic theory, the role of the 
supersymmetry is to restrict the $b\ll 1 $ and $c\ll 1$ fields to their 
ground states, up to gauge transformation.

We emphasize that, 
in formulating the transition to bosonic string theory, we have not
integrated out any fields, and no information contained in the original type 0 phase
has been lost.  The late-time description is achieved entirely through
canonical variable redefinitions.  In one set of variables
(referred to as {\it UV variables} in \cite{previous3}) the theory is
weakly-coupled in the initial phase, while in another set
(denoted {\it IR variables} in \cite{previous3}) the theory is weakly-coupled
in the late-time limit.  Since no information is lost in moving to
IR variables, in which the theory reduces to bosonic string theory at
large $X^+$, this constitutes a precise duality.  Because of the
cosmological nature of the transition (meaning that time evolution in the
target space drives RG flow on the worldsheet), and because there is again a
transfer of central charge from the UV fermions to the dilaton, we
refer to this transition as {\it c-duality}.

\ \\

\paragraph{Relationship between the type 0 web and the bosonic string}

\begin{figure}[htp]
\begin{center}
\includegraphics[width=2.8in,height=6.0in,angle=0]{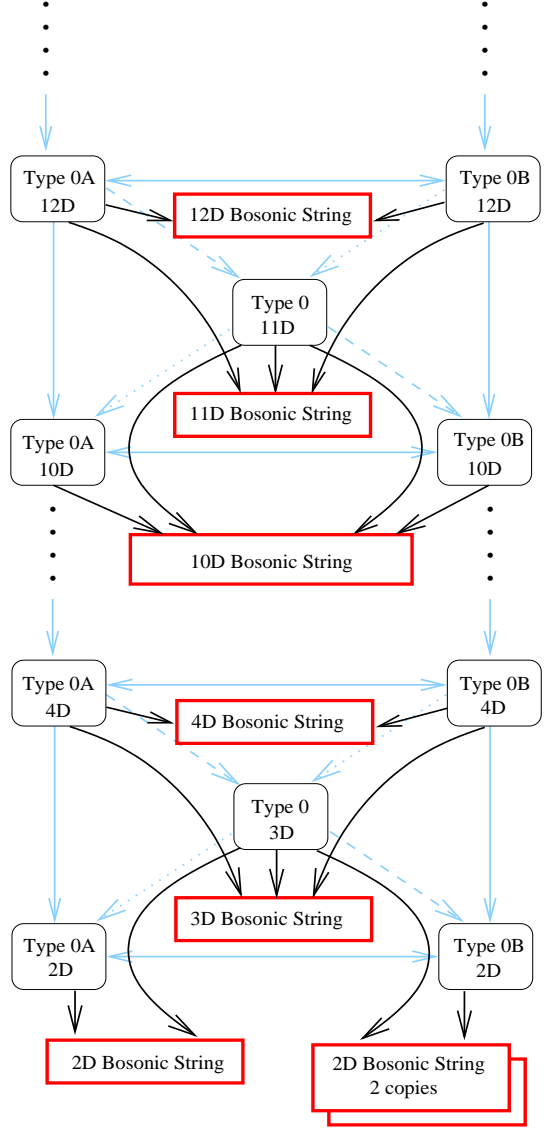}
\caption{\footnotesize
Transitions to bosonic string theory in two dimensions and higher
can occur via c-duality, starting from points in the type 0 series.
The transitions connect type 0 theories in $D$ dimensions to bosonic string 
theory in $D$ (straight, solid arrows) 
or $D-1$ (curved, solid arrows) noncompact dimensions, with compact
current algebra factors.
The straight, solid arrows are c-duality
transitions, and the curved solid arrows
are the detuned versions of tuned dimension-changing
transitions with $\Delta D = 1$.
The detuned transitins combine
c-duality and dimension quenching.  At the bottom of the type 0 series, 
c-duality connects type 0B in $2D$ to two copies of bosonic string theory
in $2D$, while the analogous flow connects type 0A to a single copy of 
bosonic string theory in $2D$.  Both of these endpoints are also 
connected by tuned transitions to type 0 string theory in $3D$.  
The transitions to bosonic 
string theory in this figure are understood to be superimposed on the
corresponding structure in the type 0 hierarchy depicted in Fig.~\ref{0fig}
(which is displayed in light blue; color is available in the electronic version). }
\label{bosefig}
\end{center}
\end{figure}

Bosonic string theory in various dimensions can be reached via c-duality from
type 0 string theory in various dimensions.  
The basic transitions induced by the tachyon profile 
${\cal T} = \m \, \exp{\b X\uu +}$
connect $D$-dimensional type 0 string theory to the bosonic 
string in $D$ noncompact
dimensions with an $SO(D-2)\ll L\times SO(D-2)\ll R$ current algebra.
The current algebra retains a memory (via the set of representations
that appear) of whether the parent type 0 theory was
type 0B or 0A.  In both
cases the current algebra contains some states in which the fermions are
all simultaneously periodic.  The GSO projection in the R sector of the type 0 parent theory
determines the state of the fermion zero modes in the sector of the current
algebra with periodic fermions.  That is, the type of RR forms that
appear in the type 0 parent theory determine the $SO(D-2)\ll L \times SO(D-2)\ll R$
chirality of the bispinor states of the current algebra.

A related set of transitions can be obtained by ``detuning'' the 
tuned dimension-changing type 0 transitions.
For instance, we can modify the transition from type 0 in $D$ dimensions
to type 0 in $D-1$ dimensions by deforming $\mu'$ away from its tuned value. 
This is equivalent to deforming the $D-1$ dimensional final state with a tachyon vev
of the form ${\cal T} = (\mu' - \m'_{\rm tuned}) \exp{\b X\uu +}$.
This induces a further transition to the bosonic string in $D-1$ noncompact dimensions
with an $SO(D-3)\ll L \times SO(D-3)\ll R$ current algebra.
We will refer to this type of solution as a {\emph{detuned}} transition.

The detuned transitions follow the trajectories of the 
tuned transitions arbitrarily closely
over an arbitrarily long time.  Ultimately they deviate and land on a final state 
described by the bosonic string.  The detuned solutions can therefore be thought of 
as hybrid transitions combining dimension quenching and c-duality,
reducing the number of noncompact dimensions {\it and} changing the type 
of string theory relative to the initial state.


These transitions are depicted in Fig.~\ref{bosefig}.
Direct transitions to bosonic final states are drawn 
with solid straight lines, while
transitions that combine c-duality with dimension quenching are shown with
solid curved lines.  The latter are meant to suggest that, in starting from
an initial type 0 phase in even dimension ($D$), the worldsheet theory
evolves along a pathway that is very similar to the tuned dimension-quenching
transition that lands on type 0 string theory in odd dimension ($D-1$).
In the large $X^+$ region, however, the flow misses the type 0 final phase
and lands instead on bosonic string theory in odd dimension ($D-1$).

\paragraph{c-duality in D=2}
\ \\
The c-duality transition in two dimensions involves various
subtleties that depend on the GSO projection.  
In type 0B there is a physical Ramond ground state 
$\kket{R}$ of weight
zero that survives the GSO projection.  Combining this with the NS ground state $\kket{NS}$,
one can construct projection operators onto states of the form
$\hh \left ( \kket{NS} \pm \kket{R} \right )$.  These
projectors break states of the theory into two sectors that are decoupled in the OPE
\cite{stacks,Takayanagi:2003sm,newhat}. 
This means that the type 0B string 
describes two decoupled universes in closed-string perturbation theory.
However, it is known that nonperturbative effects associated with
D-instantons lift the degeneracy of the projection operators and
couple the two universes \cite{Takayanagi:2003sm,newhat}.

The degeneracy of the NS and R ground states persists in the Berkovits-Vafa embedding
describing the $X\uu + \to \infty$ region of the theory.  
In other words, 
there are still projection operators at $X\uu +\to \infty$ 
on the worldsheet that decompose the string Hilbert space
into two sectors that are not coupled through perturbative interactions.  
To the extent that we can trust 
perturbation theory, the final state of the c-duality transition from type 0B
string theory is therefore two decoupled copies of $2D$ bosonic string theory in a
spacelike linear dilaton background.  It is not
clear, however, whether this description is correct nonperturbatively, or whether D-brane
effects may couple the two universes as they do in the type 0B initial background.

D-instanton contributions can be suppressed, but not without altering 
the properties of the theory that render the transition solvable.
In the presence of a Liouville wall in the initial type 0B theory, 
D-instantons are exponentially suppressed, and the same may be true for the final state.  
However, our c-duality transitions can only be constructed as exact CFTs in the
strong-coupling limit, for which the Liouville wall is absent.  (Otherwise, there 
would be a nontrivial interaction in conformal perturbation theory between insertions of 
$\exp{ \b X\uu 1}$, associated with the Liouville wall, and $\exp{\b X\uu +}$,
associated with the c-duality transition.)

In type 0A string theory in $2D$, the fundamental string has no physical Ramond states
whatsoever.  The same is true of the Berkovits-Vafa embedding of the 
$2D$ bosonic string when the GSO projection in the R sector is $\mfw = -1$.
Although the Ramond sector has four degenerate ground
states, it is not possible to reverse the GSO projection by acting with a fermion
zero mode while satisfying the physical state conditions \cite{previous3}.
Only one ground state satisfies the $G\ll 0, \tilde{G}\ll 0$ physical
state conditions, and if the corresponding Ramond state is 
removed by the GSO projection, the Ramond sector is completely empty.
In this case the CFT has only one component, 
and the endpoint of the
transition describes a single copy of bosonic string theory in two dimensions.

\section{\label{classify}A classification theorem}
It would be useful to classify all possible supercritical vacua
that can make transitions to points in the $10D$ supersymmetric 
moduli space.  As we have seen in this paper, supercritical starting
points above supersymmetric vacua are somewhat constrained, but still diverse.
In this section we will give a partial classification of the
possibilities under certain simple conditions.

\subsection{Supercritical vacua above type II in 10 dimensions}
The set of supercritical vacua that can relax to type II string
theory in 10 dimensions by classical
condensation of NS tachyons is
large.  We can narrow the scope of our classification
by restricting to cases in which the worldsheet description
of the flow can be analyzed reliably with a semiclassical treatment
of the worldsheet fields transverse to
the $X^\pm$ directions.  (Worldsheet quantum corrections
resulting from the dynamics of the $X^\pm$ directions vanish when the
tachyon gradient is lightlike \cite{previous2}.)  In other words,
we will consider cases in which the worldsheet stress tensor
and supercurrent of the initial-state background are well approximated 
quadratic expressions in free fields.  Furthermore, the
tachyonic perturbation is assumed to be well approximated everywhere
in the classical worldsheet vacuum manifold by a quadratic expression
in free fields, dressed with an exponential of the lightlike 
direction $X\uu +$.  (For paradigmatic examples and further analysis 
of this type of transition, see \cite{previous2}.)  

Let us list our technical assumptions.  In addition to
the requirement of semiclassical worldsheet dynamics, we will
restrict to cases for which:\footnote{Of course, one can 
expand the search by relaxing some of these conditions.}
\begin{itemize}
\item{The final state is a single copy of unorbifolded
ten-dimensional space with flat string-frame metric and
lightlike linear dilaton, rolling to weak coupling in the
future;}
\item{The GSO projection of the final theory is the chiral GSO
projection of the type IIA/B string;}
\item{The initial state is described by an oriented string
theory;}
\item{The continuous worldsheet gauge symmetry of the initial theory 
is linearly realized $(1,1)$ 
superconformal invariance (i.e.,~the initial theory must be a 
type 0 or type II string);}
\item{The (1,1) worldsheet supersymmetry is unbroken in the final-state vacuum manifold
(if the worldsheet supersymmetry is spontaneously broken and the vacuum energy is tuned to zero,
the resulting physics is that of the bosonic string
\cite{previous3}).}
\end{itemize}

\paragraph{Discrete gauge symmetry}
\ \\
We will first prove that the discrete worldsheet gauge symmetry in
the initial state must be $\IZ\ll 2 \times \IZ\ll 2$ acting
chirally on the supercurrents $\tilde{G}, G$.  We know that the
final-state discrete gauge symmetry is the group generated by
the operators
$(-1)\uu{F\ll{L\ll W}}$ and $(-1)\uu{F\ll{R\ll W}}$,
which act trivially on the $X\uu M$ and with
a $-1$ on $\tilde{G},G$ respectively.
The tachyon condensation is
described by semiclassical worldsheet RG flow, which can
only break, and not enhance, gauge symmetries.  The discrete 
worldsheet gauge symmetry of the initial state must therefore be
at least $\IZ\ll 2 \times \IZ\ll 2$, which acts as a chiral R-parity
on the supercurrents $\tilde{G}, G$.  

We can also show that the discrete gauge group is no larger 
than $\IZ\ll 2 \times \IZ\ll 2$.  
The key idea is that for a larger group, the worldsheet theory at $X\uu + \to \infty$
must contain multiple components as direct summands of the CFT, generated by the
mechanism described in \cite{stacks}.
If the orbifold group were larger than the minimal $\IZ\ll 2 \times \IZ\ll 2$, 
there would have to be an element $g\ll 3$ that
acts trivially on both supercurrents, but nontrivially on
matter degrees of freedom.\footnote{Here we are using the
fact that the gauged 
worldsheet supersymmetry on each side of the string is
only ${\cal N} = 1$, and not ${\cal N}=2$ or higher.}
We now focus on the sector twisted by $g\ll 3$.  
Since we have assumed the RG flow is semiclassical,
the NS tachyon describing the condensation must be untwisted. 
This twist is therefore a good symmetry throughout the entire flow, including
in the IR limit.

Now we consider the final-state theory and focus on the  
sector twisted by $g\ll 3$.  
By assumption, we have a final
state that is an unorbifolded type II theory, so $g\ll 3$ acts trivially
on all infrared degrees of freedom.  The physics
of these types of discrete gauge symmetries was analyzed in \cite{stacks}.
The NS/NS ground state in the sector
twisted by $g\ll 3$ has weight zero in the infrared.  The Hilbert space built
on the twisted vacuum gives rise to an identical copy of
the ten-dimensional universe in the untwisted vacuum.
Therefore, if the discrete
gauge group in the initial theory is larger than the minimal $\IZ\ll 2
\times \IZ\ll 2$, the final state has at least two disconnected
components.  We conclude that the GSO group of the 
supercritical theory must be exactly $\IZ\ll 2 \times \IZ\ll 2$
if the final state is critical type II string theory on
a single connected component $\IR\uu{9,1}$.

\subsection{Geometric action of the GSO group}
In addition to acting on the supercurrents, the generators
$(-1)\uu{F_{L_W}}$ and $(-1)\uu{F_{R_W}}$ of the GSO group
may act on the coordinates $X\uu M$ of the D-dimensional
initial theory.  In the general case, this geometric action must 
respect the isometries of the initial state.  In the limit where the
initial state is flat and noncompact, the geometric action of the GSO group
must be a subgroup of the Euclidean group, which means it must be a
combination of shifts and rotations.  Since the subgroup is $\IZ\ll 2\times \IZ\ll 2$,
the rotations must be $180$ degree reflections at most.

We now wish to determine how the GSO group $\IZ\ll 2\times \IZ\ll 2$ acts
on the coordinates $X\uu{0,\cdots , D-1}$.  
Functions of $X\uu M$ can be decomposed into eigenfunctions
$f\ll{\pm\pm\pr}(X)$ under the action of $\IZ\ll 2 \times \IZ\ll 2$.
It is easy to see that only $f\ll{++}$ and $f\ll{--}$ are allowed
by modular invariance, and not $f\ll{-+}$ or $f\ll{+-}$.
If we have a function $f\ll{-+}$ that
is odd under $\mflw$ and even under $\mfrw$, acting on
the normal-ordered operator $:f(X):$ with the right-moving
supercurrent $G$ gives an NS/NS state of half-integer spin surviving
the GSO projection.  This is forbidden by modular invariance.
We conclude that $(-1)\uu{F_{R_W}}$ must act identically
to $(-1)\uu{F_{L_W}}$ on the coordinates $X\uu M$, meaning that the
geometric action of $(-1)\uu{F_W}$ must be trivial on all $D$ coordinates.

Next we will show that, in addition to the total fermion parity $\mfw$, 
the {\it chiral \rm} fermion parity $\mflw$ must act trivially on the vacuum 
manifold ${\bf M}\ll{\rm IR}$ of the
{\it final \rm} configuration, 
assuming the final state is type II rather than type 0 on macroscopic scales.

\paragraph{Lemma}
\ \\
Consider the action of $(-1)\uu{F_{L_W}}$ on the zero-energy
vacuum manifold ${\bf M}\ll{\rm IR}$ of the worldsheet potential.
By assumption, ${\bf M}\ll{\rm IR}$ has ten dimensions.  
Let $L$ be the typical distance scale of ${\bf M}\ll{\rm IR}$ (where $L = \infty$
in the case for which the final state is noncompact).
Our assumption that 
the flow is semiclassical on the worldsheet implies
that $L \gg \sqrt{\apr}$.
It is therefore clear that $(-1)\uu{F_{L_W}}$ must act trivially on ${\bf M}\ll{\rm IR}$
if the local physics is to be type II, rather than type 0,
on scales smaller than $L$.

For instance, there must exist an unstable
mode of the tachyon ${\cal T}$ if the operation $\mflw$ acts
nontrivially on ${\bf M}\ll{\rm IR}$.  To see this, we can decompose functions on
${\bf M}\ll{\rm IR}$ into even modes $f\ll{++}$ and odd modes $f\ll{--}$.
The GSO projection removes the $f\ll{++}$ modes of the
tachyon and retains the $f\ll{--}$ modes.
If $\mflw$ acts nontrivially on ${\bf M}\ll{\rm IR}$, then the
odd modes $f\ll{--}$ must exist.    If 
$\L\sqd$ is the eigenvalue of the lowest odd mode $f\ll{--}$,
the frequency-squared of the lowest surviving tachyon mode is $\L\sqd - {2\over\apr}$.
The eigenvalues of the Laplacian for all
eigenmodes are of order $L\uu{-2}$ in the semiclassical limit $L \gg \sqrt{\apr}$.
In this limit, therefore, there are modes of the tachyon with imaginary frequency, which
grow exponentially in time.  

We conclude that the final 10-dimensional state cannot be supersymmetric
type II string theory unless the chiral GSO projection acts trivially on the vacuum
manifold ${\bf M}\ll{\rm IR}$.  We can use this fact to constrain
the action of the chiral GSO projection on the $D$ coordinates
of the initial supercritical state. 
We consider two possible cases: 1) No point of the initial theory is
fixed under the action of $\mflw$, or 2) there are nonempty fixed
loci under the action of $\mflw$.

\paragraph{Case 1: No fixed loci of $\mflw$}
\ \\
Suppose we adopt the first case, in which
$\mflw$ acts on the D-dimensional initial state without fixed points.
Since the 
vacuum manifold ${\bf M}\ll{\rm IR}$ is a subspace of the original
$D$-dimensional spacetime, any fixed point in ${\bf M}\ll{\rm IR}$ is necessarily
a fixed point in the $D$-dimensional space, so
$\mflw$ must act on ${\bf M}\ll{\rm IR}$ without fixed points as well.
The GSO projection means
that the physical $10D$ geometry is a fixed-point-free quotient of ${\bf M}\ll{\rm IR}$,
by the action of $\mflw$, with antiperiodic boundary conditions for ${\cal T}$ 
around the ${\IZ}\ll 2$ one-cycle.
By the lemma given above, the local physics is
that of type 0, not type II, on scales smaller than $L$.

This result is intuitively clear when one considers the necessity of having
massless fermions in $10D$.  Spacetime fermions arise only
from twisted sectors of the chiral GSO projection.  If this projection
has no fixed points, then all such twisted sectors describe stretched
strings, whose length is the typical scale $L$ of the manifold.  As long as 
$L \gg \sqrt{\apr}$, all spacetime fermions are much
heavier than the string scale if there are no fixed loci of $\mflw$.  
There are therefore no light fermions that could fill 
out supersymmetry multiplets in the 
$X\uu + \to \infty$ limit.

\paragraph{Case 2: Nonempty fixed loci of $\mflw$}
\ \\
In the second case, we consider a fixed locus of $\mflw$, whose
codimension we denote by $n\ll{\rm odd}$.  (If there are multiple disconnected
components of the fixed locus, each one may be assigned its own
codimension $n\ll{\rm odd}$.)  The vacuum manifold of the
bosonic potential must lie entirely within the fixed locus. 
Otherwise, the chiral GSO projection acts nontrivially on ${\bf M}
\ll{\rm IR}$, giving a background whose local physics
is type 0, not type II.

Let $Y\uu A$ (with $A = 1,\cdots, n\ll{\rm odd}$) be local coordinates
normal to the fixed locus.  These are, by definition,
odd under the action of $(-1)\uu{F_{L_W}}$.  
Let $X\uu \pm ,X\uu a$ ($a = 1,\cdots ,n\ll{\rm even} - 2$) be
the coordinates longitudinal to the fixed locus.
The sum $n\ll{\rm even} + n\ll{\rm odd}$ is equal to the total
dimension $D$ of the initial state.  We will now
show that $n\ll{\rm even} - n\ll{\rm odd}$ is
an invariant of any semiclassical RG flow.

The tachyon ${\cal T}$ couples to the worldsheet as a 
superpotential:
\be
\Delta {\cal L} = {i\over{2\pi}} \int d\theta\ll + d\theta\ll - 
~{\cal T}(X) \ ,
\ee
where
\be
{\cal T}(X) = \exp{\b X\uu +} f(X\uu a, Y\uu A) \ .
\ee
We wish to expand around an arbitrary point $x\uu a\ll {(0)}$ on the vacuum manifold,
choosing coordinates such that $x\uu a\ll{(0)} = 0$.
The condition that the flow be semiclassical means that the
flow is controlled by terms that are constant, linear and quadratic 
in $X,Y$ and their superpartners, with terms of cubic and higher
order making no qualitative difference to the transition.
The function $f$ can therefore be approximated as
a quadratic function of $X$ and $Y$ in an expansion around the origin.

By assumption, our vacuum manifold preserves worldsheet supersymmetry.
The $F$-term conditions
\be
\del_{X\uu +} {\cal T} (0) = \del_{X\uu a} {\cal T}  (0) = \del_ {Y\uu A} {\cal T} (0) = 0
\ee
mean that the constant and linear terms in $f$ must be zero at $X\uu a = Y\uu A = 0$.  
In other words, the function $f$ must vanish to second order at 
any point on the vacuum manifold.
The form of the quadratic term is constrained by invariance under the chiral
GSO projection. 
The superspace measure $d\tilde{\theta} d\theta$ is odd under 
$(-1)\uu{F_{L_W}}$, so the function ${\cal T}(X,Y)$ must
be odd as well, linking only $X$ coordinates and $Y$ coordinates.  We thus have
\be
{\cal T} = \exp{\b X\uu +} ~\left[ q\ll{aA} X\uu a Y\uu A + ({\rm  cubic}) \right]
\ee
around any point on the vacuum manifold, for some constant matrix $q\ll{aA}$.
It follows that
$n\ll{\rm even} - n\ll{\rm odd}$ 
is an invariant of tachyon condensation when $\apr$ corrections are
small in the dynamics of the $X\uu a, Y\uu A$ sector.
Since $n\ll{\rm odd}$ vanishes in the final, critical type II theory,
we must have $n_{\rm odd} = \hh (D-10)$ in the initial state.  
This reduces the set of possibilities to the
models described in \cite{previous2}, in the limit where the initial state has
all dimensions noncompact and flat.

\section{Other simple type 0/type II theories and connections between them}
We have classified the set of supercritical vacua that can decay
to type II string theory via classical solutions
describing the lightlike condensation of perturbative instabilities, under the assumption
that all scales $L$ transverse to the light cone $X\uu \pm$ are much larger
than the string scale: $L \gg \sqrt{\apr}$.
The full set of supercritical type II and type 0 theories is 
immensely larger, and we will not attempt to classify it here.  Instead,
we will examine a simple subset of type 0 and type II theories in
diverse dimensions, and chart the connections between them.

The sector of the supercritical landscape
we will describe consists of theories with linearly-realized local
(1,1) superconformal symmetry on the worldsheet, 
with all dimensions noncompact and
flat, and with a possible orbifold singularity at the origin.
We will assume the discrete gauge group is the minimal 
$\IZ\ll 2\times \IZ\ll 2$
GSO group; larger groups can be constructed by orbifolding the theories we
discuss here.

The worldsheet field content for such theories consists of $D$ bosons
$X\uu M$  and their worldsheet superpartners $\psi\uu M, \pst\uu M$.
By the same argument given above, $(-1)\uu{F_W}$ (which acts
as a $-1$ on both supercurrents) must act as a 
$+1$ on all $X\uu M$.  The chiral GSO 
generator $(-1)\uu{F_{L_W}}$ can act
as a $-1$ on some of the $X\uu M$, and the various 
possibilities are tightly constrained by modular invariance.

The simplest constraint comes from the level matching condition in the 
R/NS and NS/R sectors.  We let the number of coordinates $X\uu M$ that are
even (resp.~odd) under $(-1)\uu{F_{L_W}}$ be denoted by $\nev$ (resp.~$\nod$).
If we make the standard choice of GSO projection, which is NS$_+$/R$_\pm$
and R$_\pm$/NS$_+$, then level matching requires $\nev - \nod = 10 + 16K$.
We refer to this as the {\it standard series \rm} of type 0/II hierarchies.
  Each hierarchy has a particular value of $K$, and terminates in a stable
type II theory with no geometric orbifold singularities.  We can also consider the
opposite projection on the NS side of the NS/R and R/NS sectors, namely
NS$_-$/R$_\pm$ and R$_\pm$/NS$_-$.  These sectors are level-matched
if and only if $\nev - \nod = 2 + 16K$.  We denote the set of string 
theories defined this way the {\it Seiberg series. \rm}
\footnote{The series is so named because they are based on
a generalization of the recently discovered two-dimensional
type II theories \cite{seibergcircles}.  Indeed, the models of
\cite{seibergcircles}
constitute the lowest-dimensional example of the
Seiberg series.}
In all cases, the GSO projection in the NS/NS sector is
$+/+$, corresponding to a GSO projection of $-/-$ in the matter sector.

For any value of $K$ and $n\ll{\rm odd}$, $D$ is always even in both series, and
both have consistent OPEs and modular-invariant partition functions.
For the stable, bottom rungs (that is, $n\ll{odd} = 0$)
of the standard series, the partition function is 
\be
Z \equiv \int \ll{\bf F} {{d\t d\bar{\t}}\over{\t\ll 2\sqd}} |\eta(\t)|\uu{- 2(D - 2)}
~\lrdd 4\pi\sqd \apr \t\ll 2
\rrdd\uu{- {{D - 2}\over 2}} \lrdd I\ll{\rm NS+}\st(\bar{\t})  - I\ll{\rm R\pm}\st(\bar{\t})
 \rrdd \lrdd I\ll{\rm NS+}(\t)
- I\ll{\rm R\pm\pr}(\t) \rrdd \ ,
\nn
\\
\ee
where
\be
I\ll{NS\pm} (\t) &\equiv &\hh 
\lsqq (Z\uu 0{}\ll 0(\t) )\uu{{{D - 2}\over 2}} 
\mp (Z\uu 0{}\ll 1(\t) )\uu{{{D - 2}\over 2}} \rsqq \ ,
\nn\\
&&
\nn\\
I\ll{R +} (\t) &\equiv& I\ll{R - } (\t) \equiv 
 \hh (Z\uu 1{}\ll 0(\t))\uu{{{D - 2}\over 2} } \ .
\ee
with
\be
Z\uu\a{}\ll\b (\t) \equiv {1\over{\eta(\t)}} \th\ll{\a\b} (0,\t) 
\ee

For the stable, bottom rungs of the Seiberg hierarchies, the partition function is 
\be
Z &\equiv& \int \ll{\bf F} {{d\t d\bar{\t}}\over{\t\ll 2\sqd}} |\eta(\t)|\uu{- 2(D - 2)}
~\lrdd 4\pi\sqd \apr \t\ll 2 
\rrdd\uu{- {{D - 2}\over 2}} 
\nn\\
&&
\nn\\
&&
\times \lrdd I\ll{NS+}(\bar{\t})\st I\ll{NS+}(\t) 
- I\ll{\rm NS-} \st(\bar{\t})  I\ll{R\pm\pr}(\t) - I\ll{R\pm}(\bar{\t}) I\ll{NS-} (\t)
+ I\ll{R\pm}(\bar{\t}) I\ll{R\pm\pr}(\t)
  \rrdd \ .
\nn\\
&&
\ee

For $\nod = 0$, the theories are tachyon
free, since the matter part $V\ll{(\hh,\hh)}$ of the vertex operator must
have GSO charges $-/-$, but all embedding coordinates are even under $\mflw$.
For $\nod \neq 0$, there are always tachyons.  If $Y\uu A$ are the odd
coordinates and $X\uu a$ are the even coordinates, any function
$f(Y,X)$ satisfying $f(-Y,X) = - f(Y,X)$ corresponds to the matter
component of an allowed
tachyon vertex operator $:f(X,Y):$.  Solvable classical backgrounds exist
describing tachyon condensation that reduce the dimension $D$ while keeping
$\nev - \nod$ constant.  Such solutions are described by
tachyon profiles of the form
\be
{\cal T} = \m~\exp{\b X\uu +} \sum\ll
{a = 1}\uu{\Delta \nod} X\uu {a+1} Y\uu a \ .
\ee
By the mechanism of central charge transfer described in \cite{previous2}, 
the solution transfers ${3\over 2} \Delta D = 3 \Delta\nod$ units of central charge to
the dilaton gradient, by an exactly calculable 
renormalization effect on the worldsheet.

This generates an infinite number of
hierarchies of cosmological string theory, parametrized by
an integer $K$.  Each hierarchy is connected
vertically by solvable dynamical transitions that preserve $K$ and reduce
$\nod$ and $\nev$ by the same amount.  The lowest element of
each hierarchy, with $\nod = 0$, is tachyon-free.

These hierarchies are also connected
`laterally' to the hierarchy of unorbifolded type 0 strings.
The horizontal connections
can be understood either as orbifolding or 
as T-duality on a twisted circle. 
Starting from type 0 in $D$ dimensions, one can orbifold
by reflection of $\nod$ directions, together with the action of
$\mflw$, to reach a consistent, modular-invariant theory
when $D - 2 \nod$ is equal to $10 + 16 K$ or $2 + 16 K$.  This
orbifolding produces a theory with $\nod$ odd directions in the hierarchy
labeled by $K$.  Starting with one of the type II theories 
or type 0 orbifolds,
one can orbifold by spacetime fermion
number mod two ($\mfs$) and reach an unorbifolded type 0 phase.  
(Orbifolding by $\mfs$ 
removes the geometric singularity of the type 0 orbifolds
along with the twisted states that live there, since those twisted sectors are
all spacetime fermions.)

Instead of orbifolding, one can also move horizontally between type 0 and
the chiral-GSO hierarchies by a discrete Wilson line construction, as 
described in \cite{wilsonization}.  We can start with 
type 0A/B string theory,
compactifying on a circle with a Wilson line for the symmetry $g$.
Taking the radius to zero and transforming to T-dual variables
produces the same effect as orbifolding the type 0B/A theory by
$g$, if such an orbifold is modular-invariant.  (If the orbifold is not
modular-invariant, the T-dual description of the small-radius limit
is again an unorbifolded type 0 theory.)  To avoid switching type 0A with
type 0B, one may combine $g$ with $\mfls$ (which reverses the signs of all
R/R sectors) in the action implemented by the Wilson line.  To move from the
chiral-GSO hierarchies to the type 0 hierarchy, one may compactify on a circle with
Scherk-Schwarz 
boundary conditions for the spacetime fermions and take the
small-radius limit, switching to T-dual variables.

\paragraph{Partition functions for $n\ll{\rm odd}\neq 0$}
\ \\
The tachyon-free case $n\ll{\rm odd} = 0$ is the simplest, in that
all bosons $X\uu M$ make the same contribution to the partition 
function.  For $n\ll{\rm odd} \neq 0$, we have contributions from odd (orbifolded)
coordinates $Y\uu A$.
To organize these contributions, we define
\be
J\uu{\tilde{\a} |\a}{}\ll{\tilde{\b} |\b} (\t, \bar{\t})
\ee
to be the path integral for all bosons and fermions
in the sector with boundary conditions defined by the four
periodicities $\a,\tilde{\a},\b,\tilde{\b}$ for the two
supercurrents $G,\tilde{G}$ on the two-cycles.  That is,
a zero represents an antiperiodic supercurrent and
a one represents a periodic supercurrent.  By our choice of 
GSO projection, the perodicity of the odd bosons $Y\uu A$ is
correlated with that of the supercurrents.  In particular, an
odd boson $Y$ comes back to itself
with a phase of $(-1)\uu{\a + \tilde{\a}}$
around a cycle when the supercurrents have
periodicities $\a,\tilde{\a}$.  
The path integrals ${\bf Y}\uu \a{}\ll\b$ for an odd boson are
\be
{\bf Y}\uu 0{}\ll 0 (\t,\tb) & \equiv &
\lrdd 4\pi\sqd\apr\t\ll 2\rrdd\uu{-\hh}
|\eta(\t)|\uu{-2} \ ,
\nn\\
&& 
\nn\\
{\bf Y}\uu 0{}\ll 1 (\t,\tb) &\equiv & {{|\eta(\t) |}\over
{|\theta\ll{10}(0,\t)|}}   \ ,
\nn\\
&&
\nn\\
{\bf Y}\uu 1{}\ll 0 (\t,\tb) &\equiv & {{|\eta(\t) |}\over
{|\theta\ll{01}(0,\t)|}}   \ ,
\nn\\
&&
\nn\\
{\bf Y}\uu 1{}\ll 1 (\t,\tb) &\equiv &  {{|\eta(\t) |}\over
{|\theta\ll{00}(0,\t)|}} \ . 
\ee
The path integrals for the odd fermions are the same
as the $Z\uu \a{}\ll\b$, except that the right-moving
fermions have the periodicities of the left-moving
supercurrents, and vice-versa.  So the path integral for
the bosons and fermions in the Y multiplets is equal to
\be
J\uu{\tilde{\a},\a}\ll{~\tilde{\b},\b} (\t,\tb)
\equiv \lsqq {\bf Y}\uu{\a + \tilde{\a}}\ll{~\b + \tilde{\b}}
(\t,\tb) 
\lrdd Z\uu{\a*}{}\ll\b (\tb) 
 Z\uu{\tilde{\a}}{}\ll
{\tilde{\b}} (\t) \rrdd\uu\hh   \rsqq \uu{n\ll{\rm odd}} \ ,
\ee 
when $G$ has the boundary conditions $\a,\b$ and
$\tilde{G}$ has boundary conditions $\tilde{\a},\tilde{\b}$.
We can include the fermionic superpartners of the even 
coordinates as well, which always have periodicities that
match those of the supercurrents.  
The periodicities of the superghosts $\b,\g,\tilde{\b},\tilde{\g}$
match those of the supercurrents as well, and exactly
cancel the path integral of two real fermions of each chirality.

Defining 
$I\uu{\tilde{\a},\a}\ll{~\tilde{\b},\b} (\t,\tb)$  to be
the path integral over the $Y\uu A$ degrees of freedom
and their superpartners, the fermions in the
$X\uu a$ multiplets, and the superghosts, we find
\be
I\uu{\tilde{\a},\a}\ll{~\tilde{\b},\b} (\t,\tb)
= J\uu{\tilde{\a},\a}\ll{~\tilde{\b},\b} (\t,\tb)~
\lrdd Z\uu\a{}\ll\b (\t) Z\uu{\tilde{\a}*}{}\ll{\tilde{\b}}
(\bar{\t}) \rrdd\uu{{{n\ll{\rm even} - 2}\over 2}} \ .
\ee
From these path integrals, we define partition functions
over the $Y,\psi,\pst$ and superghost degrees of freedom in each
sector:
\be
I\ll{\rm NS+/NS+} &\equiv& {1\over 4} \lrdd
I\uu{0,0}\ll{~0,0} - I\uu{0,0}\ll{~1,0}
- I\uu{0,0}\ll{~0,1} + I\uu{0,0}\ll{~1,1} \rrdd  \ ,
\nn\\
&&
\nn\\
I\ll{\rm NS\pm\pr /R\pm} & \equiv & {1\over 4} \lrdd
I\uu{0,1}\ll{~0,0} \mp\pr I\uu{0,1}\ll{~1,0} 
 \rrdd  \ ,
\nn\\
&&
\nn\\
I\ll{\rm R\pm\pr/NS\pm} &\equiv & {1\over 4} \lrdd
I\uu{1,0}\ll{~0,0} \mp I\uu{1,0}\ll{~0,1}
 \rrdd \ ,
\nn\\
&&
\nn\\
I\ll{\rm R\pm\pr/R\pm} &\equiv & {1\over 4} 
I\uu{1,1}\ll{~0,0} \ ,
\ee
where we have suppressed the arguments $\t,\tb$ and used
the fact that $I\uu{1,\a}\ll{~1,\b} = I\uu{\tilde{\a},1}
\ll{~\tilde{\b},1} = 0$.
The partition functions for the $X$ and reparametrization
ghosts are the same in every sector, giving a 
total of
\be
 Z \equiv \int \ll{\bf F} {{d\t d\bar{\t}}\over{\t\ll 2\sqd}} |\eta(\t)|\uu{- 2(n\ll{\rm even} - 2)}
~\lrdd 4\pi\sqd \apr \t\ll 2
\rrdd\uu{- {{n\ll{\rm even} - 2}\over 2}}
\nn\\ \nn\\
\times \lrdd I\ll{\rm NS+/NS+} - I\ll{\rm R\pm/NS+}
- I\ll{\rm NS+/R\pm\pr} + I\ll{\rm R\pm/R\pm\pr} \rrdd
\ee
for the standard series.  For the Seiberg series, we have
\be
 Z \equiv \int \ll{\bf F} {{d\t d\bar{\t}}\over{\t\ll 2\sqd}} |\eta(\t)|\uu{- 2(n\ll{\rm even} - 2)}
~\lrdd 4\pi\sqd \apr \t\ll 2
\rrdd\uu{- {{n\ll{\rm even} - 2}\over 2}}
\nn \\ \nn \\
\times 
\lrdd I\ll{\rm NS+/NS+} - I\ll{\rm R\pm/NS-}
- I\ll{\rm NS-/R\pm\pr} + I\ll{\rm R\pm/R\pm\pr} \rrdd  \ .
\ee

\section{\label{conc}Conclusions}
The solutions described in this article
interpolate between theories that were previously thought to 
be completely distinct.  These solutions connect string theories
in different numbers of spacetime 
dimensions,\footnote{For recent developments along these lines, see
\cite{them}.} and with different
amounts of worldsheet supersymmetry.  We have also found examples
that connect superstring theories to purely bosonic string theory,
which sheds new insight on the relationship between the two theories. 
These interpolations are cosmological in nature, in that
tachyon perturbations can be interpreted as the nucleation of
domain walls moving at the speed of light, and the late-time physics is described by
the theory deep inside the tachyon condensate.  
These solutions serve to connect a much wider class
of string backgrounds to the supersymmetric duality web of
superstring theory.

\section*{Acknowledgments}
S.H.~is the D.~E.~Shaw \& Co.,~L.~P.~Member
at the Institute for Advanced Study.
S.H.~is also supported by U.S.~Department of Energy grant 
DE-FG02-90ER40542. 
I.S.~is supported as the Marvin L.~Goldberger Member
at the Institute for Advanced Study, and
by U.S.~National Science Foundation grant PHY-0503584.
The authors gratefully acknowledge discussions with
Daniel Green, John M$^{\rm{\underline{c}}}$Greevy, Albion Lawrence,
David Morrison, Eva Silverstein, Washington Taylor,
and Barton Zwiebach.

\bibliographystyle{utcaps}
\bibliography{dimchange}

\end{document}